\documentclass[prb,twocolumn,showpacs]{revtex4}
\input psfig.sty

\def\k{{\bf k}}

\def\q{{\bf q}}

\def\r{{\bf r}}

\def\v{{\bf v}}

\def\B{{\bf B}}
\def\J{{\bf J}}
\def\F{{\bf F}}
\def\E{{\bf E}}

\def\F{{\cal F}}

\def\T{{\cal T}}

\def\1v{{\bf v}}
\def\dx2-y2{d_{x^2-y^2}}

\def\ltsim{\vbox {\hbox{\lower 1.\baselineskip \hbox{$<$}} \break
                 \hbox{\lower 0.2\baselineskip \hbox{$\sim$}} } }

\begin{document}

\draft

                                 
\title{Thermal Hall conductivity of marginal Fermi liquids 
subject to out-of-plane impurities in high-$T_c$ cuprates}

\author{Mei-Rong Li}
\affiliation{Department of Physics, University of Guelph, Guelph, Ontario,
Canada N1G 2W1}

\begin{abstract}
The effect of out-of-plane impurities on the thermal Hall conductivity 
$\kappa_{xy}$ of in-plane marginal-Fermi-liquid (MFL) quasiparticles
in high-$T_c$ cuprates is examined by following the work on 
electrical Hall conductivity $\sigma_{xy}$ by Varma and Abraham 
[Phys. Rev. Lett. 86, 4652 (2001)]. It is shown that the effective 
Lorentz force exerted by these impurities is a weak function of energies 
of the MFL quasiparticles, resulting in nearly the same temperature 
dependence of $\kappa_{xy}/T$ and $\sigma_{xy}$, indicative of 
obedience of the Wiedemann-Franz law. The inconsistency of the 
theoretical result with the experimental one is speculated to be 
the consequence of the different amounts of out-of-plane impurities in 
the two YBaCuO samples used for the $\kappa_{xy}$ and $\sigma_{xy}$ 
measurements.
\end{abstract}
\maketitle

\section{INTRODUCTION}

The temperature dependence of the in-plane longitudinal electrical 
conductivity $\sigma_{xx}\sim \tau_{tr}^{-1} \sim T^{-1}$ 
($\tau_{tr}^{-1}$ the transport scattering rate), being widely 
observed in optimally doped high-$T_c$ cuprates, can be understood 
based on the marginal-Fermi-liquid (MFL) hypothesis \cite{MFL1}. 
In this phenomenological theory, the MFL behavior of quasiparticles 
follows from the interaction between electrons and scale-invariant 
particle-hole excitation fluctuations. In the presence of a magnetic 
field, quasiparticles trajectories are bent by the magnetic Lorentz 
force, resulting in a Hall conductivity $\sigma_{xy}\sim \tau_{tr}^{-2} 
\sim T^{-2}$ according to the standard Boltzmann theory. This on its 
own is inconsistent with the experimental evidence $\sigma_{xy}\sim 
T^{-3}$ observed in an optimally doped YBaCuO sample \cite{ong1}.
On the other hand, the thermal Hall conductivity $\kappa_{xy}$ 
measured in {\it another} YBaCuO sample shows a $T^{-1.2}$ dependence 
\cite{ong2}, similar to an earlier experimental observation 
\cite{ong3}. The resulting Hall-Lorenz number 
$L_{xy}=e^2 \kappa_{xy}/(T\sigma_{xy})\sim T^{0.8}$ indicates 
a violation of the Hal-- Wiedemann-Franz law \cite{ong2}.

\vskip .2cm
Very recently, Varma and Abraham (VA) \cite{va1} proposed an intriguing 
explanation for the unusual $\sigma_{xy}$ in terms of MFL 
quasiparticles subject to out-of-plane impurities which sit 
between conducting CuO$_2$ planes, most likely the deficient oxygens 
in CuO chains. VA argued \cite{va2} that the very existence of these 
impurities and their peculiar effects on the in-plane quasiparticle 
lifetime can be verified from the experimental angle-resolved photoemission 
spectroscopy (ARPES) data \cite{Valla}.
The out-of-plane impurities differ from in-plane impurities by the fact 
that the former are forward scatterers and have negligible effect on 
$\sigma_{xx}$. However, in the presence of a magnetic field, they
impose an effective Lorentz force on the quasiparticles when the 
quasiparticles have anisotropic scattering rates. As a result, in the 
presence of a magnetic field, a nontrivial contribution to $\sigma_{xy}$ 
from the out-of-plane impurities is found, and shown to be proportional
to $\tau_{tr}^{-3} \sim T^{-3}$, in addition to the customary $T^{-2}$ 
term, 
\begin{eqnarray}
\sigma_{xy} = e^2 (a_c\, T^{-2}+b^{({\rm imp})}_c\, T^{-3}),  
\label{hall}
\end{eqnarray}
where $a_c$ and $b^{({\rm imp})}_c$, found in Eq. (\ref{abc}) below, 
are dependent on the detailed band structure, and $b^{({\rm imp})}_c$ 
also depends on the out-of-plane impurity density and the angle 
dependence of the scattering from the impurities. For the YBaCuO sample 
used in the experiments and in the temperature regimes where the 
measurements were made \cite{ong1}, the impurity term dominates over 
the customary term, leading to the observation $\sigma_{xy}\sim T^{-3}$ 
\cite{va1}. Remarkably, this simple elegant idea was further shown 
to successfully explain the complex Hall conductivity without any free 
parameters \cite{drew}. 

\vskip .2cm
In this paper we address the temperature behavior of $\kappa_{xy}$ for 
the MFL quasiparticles under the scattering of the out-of-plane 
impurities. It is of particular interest to examine the Wiedemann-Franz 
law, which is only obeyed when the electrical heat and charge currents 
decay in the same way. In an ordinary metal, this is the case when an 
elastic scattering dominates. The presence of a momentum-dependent inelastic 
scattering may change this conclusion, as a forward inelastic scattering 
process may decay the heat current more strongly than it does the 
electrical current. Such a violation of the Wiedemann-Franz law can 
usually be understood to be in association with the existence of (an) 
intrinsic energy scale(s) in the inelastic scattering. A typical example 
lies in electron-phonon interaction in the case of which the 
Wiedemann-Franz law is violated for $T<\theta_D$ \cite{ziman}, with 
$\theta_D$ the Debye temperature serving as the intrinsic energy scale.
In the present problem, the remarkable property of the inelastic 
scattering leading to the MFL behavior is the absence of such an 
intrinsic energy scale. This along with the assumption that the 
normal-state band-structure density of states $N(\theta)$ and group 
velocity $\v_\k$ are {\it constant} in energy leads to our following 
main finding in this paper. Like $\sigma_{xy}$, $\kappa_{xy}$ acquires 
both customary and unusual impurity terms,
\begin{eqnarray}
\kappa_{xy}= a_h T^{-1}+b^{({\rm imp})}_h T^{-2},  \label{hallk}
\end{eqnarray}
where $a_h$ and $b^{({\rm imp})}_h$, shown in Eq. (\ref{abh}) below,
scale with $a_c$ and $b^{({\rm imp})}_c$ in such a way that the ratio 
$b^{({\rm imp})}_h/a_h$ is the same order of $b^{({\rm imp})}_c/a_c$, 
making the Hall-Lorenz number $L_{xy}$ a weak function of $T$.
It is interesting to note that a $T$-independent $L_{xy}$ was also 
predicted by Coleman {\it et al.} \cite{Coleman} within a different 
phenomenological transport theory based on Anderson's proposal of 
two transport relaxation times \cite{Anderson}.

\vskip .2cm
The above result seems to disagree with the experiment \cite{ong2}.
It is important to note that $\sigma_{xy}$ and $\kappa_{xy}$ reported 
in Refs. \cite{ong1} and \cite{ong2} were measured in {\it different} 
YBaCuO samples. 
The sample used in the $\kappa_{xy}$ measurement is slightly overdoped, 
and has less deficient oxygens in the CuO chains. So we may expect a 
less important impurity term than the customary term in $\kappa_{xy}$. 
We suggest to reexamine $\sigma_{xy}$ in this slightly overdoped
sample to see the customary behavior of $\sigma_{xy}$.

\section{Boltzmann equation and collision integral}
We present a simple calculation leading to the aforementioned 
equation [\ref{hallk}) (as well as Eq. (\ref{hall})]. We start with the 
Boltzmann equation determining the Fermi-function deviation 
$g_\k=f_\k-f^0_\k$ in the presence of a driving force $\F$ and a 
magnetic field $\B$, 
\begin{eqnarray}
\big(\v_\k\cdot \, \F\big) \; \partial_{\xi_\k} f^0_\k
+ {e\over \hbar c} \, (\v_\k\times \B)\cdot \nabla_\k \, g_\k 
= C_\k \,,  \label{boltzmann}
\end{eqnarray}
where $\xi_\k$ is the normal-state dispersion relation being measured
from the Fermi surface (FS), and $C_\k$ the collision integral
\begin{eqnarray}
C_\k=- g_\k\tau^{-1}(\k) + \int {d\k'\over (2\pi)^2} 
C(\k,\k') g_{\k'}, \label{collision0} 
\end{eqnarray}
with 
\begin{eqnarray}
\tau^{-1}(\k)=\int {d^2 \k'\over (2\pi)^2}\, C(\k',\k)   
\label{tau0}
\end{eqnarray}
the ARPES scattering rate, and $C(\k,\k')$ encoding all information 
about microscopic interactions. The driving force in Eq. 
(\ref{boltzmann}) is $\F=-e\E$ if it is due to an electric field, and 
$\F=(-\nabla_\r T)\, \xi_\k/T$ if it is due to a temperature derivative.   

\vskip 0.2cm
Following Ref. \cite{va1}, we assume that in optimally doped 
cuprates the electrons are dominantly scattered from the unusual 
bosonic particle-hole collective modes, which results in a MFL 
behavior, and from the out-of-plane impurities. These two scattering 
processes are approximately additive,
\begin{eqnarray}
C(\k,\k')& \simeq &C_{\rm MFL}(\k,\k')+C_{\rm imp}(\k,\k'),  
\label{collision} 
\end{eqnarray}
where 
\begin{eqnarray}
&&C_{\rm imp}(\k,\k') = 2\pi \delta(\xi_\k-\xi_{\k'}) \, 
|u_i(\theta,\theta')|^2 \label{celas}
\end{eqnarray}
with $\theta$ and $\theta'$ azimuthal angles of $\k$ and $\k'$, 
respectively, and $u_i(\theta,\theta')$ sharply peaked at 
$\theta=\theta'$. By making an analogy of the bosonic collective 
modes to phonons \cite{MFL2,smith}, $C_{\rm MFL}(\k,\k')$ in Eq. 
(\ref{collision}) can be written as  
\begin{eqnarray}
&& C_{\rm MFL}(\k,\k') = C_{\rm MFL}(\xi_\k,\xi_{\k'})  \nonumber \\
&& \; = 4\pi\gamma^2_0 \, {\rm Im}\,\chi(\xi_\k-\xi_{\k'}) 
\bigg[1- f^0_{\k'}+n^0(\xi_\k-\xi_{\k'}) \bigg], 
\end{eqnarray}
where $n^0(x)$ is the boson distribution function, $\gamma_0$ is the 
coupling constant, and Im$\chi(\xi_\k-\xi_{\k'})$ is the scale-invariant 
bosonic spectral function \cite{MFL1},
\begin{eqnarray}
{\rm Im}\,\chi(\q,\omega)&=& \bigg\{ 
\begin{array}{ll}
- N_0 (\omega/T) & \;\;{\rm for} \;\; |\omega|\ll T   \\
- N_0 \, {\rm sgn} \, \omega & \;\;{\rm for} \;\; T\ll |\omega| \ll 
\omega_c,  
\end{array} 
\end{eqnarray}
with $N_0=\langle N(\theta)\rangle_{\rm FS}$ ($\langle \cdots
\rangle_{\rm FS}$ is the average over the azimuthal angle of the 
Fermi wave vector $\k_F$), and $\omega_c$ a high frequency cutoff. 

\vskip .2cm
Inserting Eq. (\ref{collision}) into Eq. (\ref{tau0}) yields the 
quasiparticle scattering rate consistent with the ARPES experiments, 
\cite{Valla}
\begin{eqnarray}
\tau^{-1}(\k)&=&\tau^{-1}_{\rm MFL}(\xi_\k,T) + 
\tau^{-1}_{\rm imp}(\theta),        \label{tau}
\end{eqnarray}
where
\begin{eqnarray}
&& \tau^{-1}_{\rm MFL}(\xi_\k,T)
\simeq \lambda_0 \, T\, \bigg\{0.97 - {\rm ln} \, 
\bigg [\tilde{f}\bigg(1+{\xi_\k\over T}\bigg)  \nonumber \\
&& \;\;\;\;\;\;\;\; \times \tilde{f}
\bigg(1-{\xi_\k\over T}\bigg)\bigg ] 
+\int^1_{-1} dy \, y \, \tilde{f}
\bigg({\xi_\k\over T}+y\bigg) \bigg\} , 
\label{MFLtau} \\
&& \tau^{-1}_{\rm imp}(\theta)
\simeq \theta_c |u_i(\theta,\theta)|^2 N(\theta) \nonumber \\
&& \;\;\;\;\;\;\;\; 
+\bigg( {\theta_c^3\over 24}\bigg) \big\{\partial^2_{\theta'}  
\big[|u_i(\theta,\theta')|^2 N(\theta') \big]\big\}_{\theta'= \theta},
\end{eqnarray}
with $\lambda_0=4\pi\gamma^2_0N^2_0$, $\tilde{f}(x)=1/(e^x+1)$ and 
$\theta_c$ a typical small angle being order of $(k_Fd)^{-1}$ ($d$
the distance of impurities to the conducting plane) characterizing the 
forward scattering nature. For $\xi_\k\ltsim T$, 
$\tau^{-1}_{\rm MFL}(\xi_\k,T)$ shown in Eq. (\ref{MFLtau}) is a weak
function of $\xi_\k$.

\vskip 0.2cm
For a small ${\cal F}$ and $B$, one finds the formal solution for 
$g_\k$ according to Eqs. (\ref{boltzmann}) and (\ref{collision0}),
\begin{eqnarray}
&& g_\k=g_{1\k}+g_{2\k},  \label{gk}\\ 
&& g_{1\k}= \int {d^2\k' \over (2\pi)^2} \T(\k,\k')
\big[ (\v_{\k'}\cdot \F) \, \partial_{\xi_{\k'}} f^0_{\k'} \big] ,  
\label{g1k} \\
&& g_{2\k}= - \int {d^2\k' \over (2\pi)^2}
\big[\hat{\T} (\v\times \B \cdot \nabla_\k)
\hat{\T}\big]_{\k,\k'}     \nonumber \\
&& \;\;\;\;\;\;\;\;\;\;\;\; 
\times \big[ (\v_{\k'}\cdot \F) \, 
\partial_{\xi_{\k'}} f^0_{\k'} \big] ,     \label{g2k}
\end{eqnarray} 
where $\T(\k,\k')$ is the matrix element of $\hat{\T}$ satisfying 
\begin{eqnarray}
\hat{\T}^{-1}=\hat{\tau}^{-1}+\hat{C}=\hat{\T}^{-1}_{\rm MFL} 
+ \hat{\T}^{-1}_{\rm imp}, \label{tt}
\end{eqnarray} 
with $\hat{\tau}$ and $\hat{C}$ having matrix elements 
$\tau(\k)\delta(\k-\k')$ and $C(\k,\k')$, respectively. Plugging 
Eqs. (\ref{gk})-(\ref{g2k}) into the expressions for the electrical and 
heat currents  
\begin{eqnarray}
\J_c=e \int {d^2\k \over (2\pi)^2} \, \v_\k \, g_\k,  
\label{chargecurrent}\\
\J_Q= \int {d^2\k \over (2\pi)^2} \,\v_\k \, \xi_\k \, g_\k,
\label{heatcurrent}  
\end{eqnarray} 
one sees clearly that $g_{1\k}$ and $g_{2\k}$ determine longitudinal 
and transverse currents, respectively.

\section{Electrical conductivities}
The electrical conductivities were already examined in detail in 
Ref. \cite{va1}, which we first review. $\partial_{\xi_{\k'}} f^0_{\k'}$ 
in Eqs. (\ref{g1k}) and (\ref{g2k}) causes the main contribution from 
the FS. A scrutiny of Eq.(\ref{tt}) leads to the important 
equation
\begin{eqnarray}
\hat{\T}\, \v_\k = \tau_{tr}(\k)  \v_\k \, +  \tau^2_{tr}(\k)\, 
\tau^{-1}_p(\theta)\; \big(\v_\k\times \hat{z}\big),  \label{lorentz}
\end{eqnarray}
where 
\begin{eqnarray}
&& \tau_{tr}^{-1}(\k)=\tau^{-1}_{\rm MFL}(0,T) + b^2(\theta) 
\tau^{-1}_u(\theta)   \nonumber \\
&& \;\;\;\;\;\;\;\;\;\;\;\;\; \simeq \tau^{-1}_{\rm MFL}(0,T),   
\label{taut} \\ 
&& \tau_p^{-1}(\theta) = 2\, b(\theta) \; \partial_\theta \, 
\tau^{-1}_u(\theta)+ [\partial_\theta b(\theta)]\,\tau^{-1}_u(\theta) , 
\end{eqnarray}
with $b(\theta)$ arising from $\partial_\theta v_{\k\alpha}
=-\epsilon_{\alpha\beta} \,b(\theta)\, v_{\k\beta}$, $\alpha, \beta=x,y,$ 
and $\tau^{-1}_u(\theta)= (\theta_c^2/ 24) \tau^{-1}_{\rm imp}(\theta)$. 
In Eq. (\ref{taut}), $\tau_{tr}^{-1}(\k)$ is the transport scattering 
rate and the contribution from the out-of-plane impurities $b^2(\theta) 
\tau^{-1}_u(\theta)$ is negligibly small due to the forward scattering 
nature. However, the second term in Eq. (\ref{lorentz}) contributed 
by the same impurities turns out to be highly nontrivial. It can be 
interpreted as an effective Lorentz force exerted by the impurities 
on the quasiparticles. Inserting Eqs. (\ref{gk})-(\ref{g2k}) into Eq. 
(\ref{chargecurrent}) results in \cite{va1}
\begin{eqnarray}
&& \sigma_{xx}(T) \simeq e^2 \, \langle N(\theta) v^2_{\k x} 
\rangle_{\rm FS} \, \tau_{\rm MFL}(0,T)  \nonumber \\
&& \;\;\;\;\;\;\;\; \;\;\;\;\;
\simeq 0.29 \, e^2 \, 
\langle N(\theta) v^2_{\k x} \rangle_{\rm FS}\, T^{-1},    
\label{sigmaxx} \\
&& \sigma_{xy}(T) \simeq  {e^3B\over mc} \bigg[\, \bar{v}_2 \; 
\tau^{2}_{\rm MFL}(0,T)   \nonumber \\
&& \;\;\;\;\;\;\;\; \;\;\;\;\;\;\;\;\;\; + \bar{v}_{imp} \; 
\tau^{3}_{\rm MFL}(0,T) \bigg]   \\
&& \;\;\;\;\;\;\;\;\;\simeq {e^3B\over mc} \bigg\{ 0.085 \, \lambda^{-2}_0 \, 
\bar{v}_2 \; T^{-2}   \nonumber \\
&&\;\;\;\;\;\;\;\;\;\;\;\;\;\;\;\;\; \;\;\;\;\; 
+ 0.025 \, \lambda^{-3}_0 \, \bar{v}_{imp}\; T^{-3} \bigg\},   \label{sigmaxy} 
\end{eqnarray}
where 
\begin{eqnarray}
&& \bar{v}_2 = \langle  N(\theta) b(\theta) \; v^2_{\k x}
\rangle_{\rm FS}, \\
&& \bar{v}_{\rm imp} = \langle N(\theta) \; \big[\partial_\theta \, 
\tau^{-1}_p(\theta)\big]\,  v^2_{\k y} \rangle_{\rm FS}. 
\end{eqnarray}
Comparing Eqs. (\ref{hall}) and (\ref{sigmaxy}) we see that
\begin{eqnarray}
&& a_c={eB\over mc} \, 0.085 \, \lambda^{-2}_0 \, \bar{v}_2,
\nonumber \\
&& b_c^{({\rm imp})}= {eB\over mc} \, 0.025 \, \lambda^{-3}_0 \, 
\bar{v}_{imp}. \label{abc}
\end{eqnarray}
It is clear that both $a_c$ and $b_c^{({\rm imp})}$ depend on the 
detailed band structure, and $b_c^{({\rm imp})}$ depends on the 
impurity scattering. As pointed out in Ref. \cite{va1} the first term 
on the right-hand-side of Eq. (\ref{sigmaxy}), the customary term, requires 
particle-hole {\it asymmetric} effect along the FS which is small 
for the high-$T_c$ cuprates. This fact brings about an interesting 
competition between the customary term and the unusual impurity term. 
For the experimentally used YBaCuO sample, the impurity term was 
analyzed \cite{va1} to dominate over the customary term leading to 
the experimental observation $\sigma_{xy}\sim T^{-3}$. 

\begin{figure}[h]
\begin{picture}(250,200)
\leavevmode\centering\includegraphics{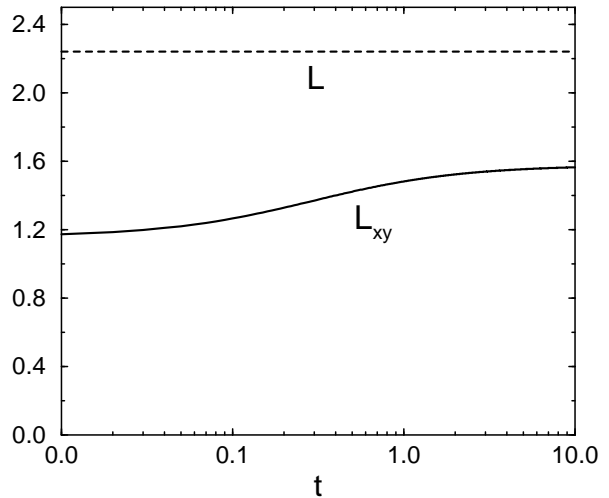}
\end{picture}
\caption{Lorenz and Lorenz Hall number as functions of
$t=T \lambda_0 \bar{v}_2/\bar{v}_{\rm imp}$.}
\label{lxy}
\end{figure}

\section{Thermal and Thermal Hall conductivities}
In calculating the heat current $\J_Q$ we have to keep the full energy 
dependence of the integrands of $g_{1\k}$ and $g_{2\k}$ in Eqs. 
(\ref{g1k}) and (\ref{g2k}). The crucial point about the MFL collective 
modes lies in their scale invariance and thus $\tau_{\rm MFL}$ and 
$C_{MFL}$ depend only on energy. Whereas in the cuprates both $v_\k$ 
and $N(\theta)$ can be reasonably assumed to be constant in energy 
near the FS (accurate up to order $T/E_F$). This means that $v_\k$ and 
$\xi_\k$ can be treated as independent variables, and 
$\hat{\T}_{\rm MFL}$ and $\hat{\T}_{\rm imp}$ in Eq. (\ref{tt}) act 
on $\xi_\k$ and $\v_\k$, respectively. Therefore, the effective Lorentz 
force exerted by the out-of-plane impurities on the MFL quasiparticles 
is a weak function of the quasiparticle energies, leading to the following 
equation similar to Eq. (\ref{lorentz}),
\begin{eqnarray}
&& \hat{\T}\, \v_\k \partial_{\xi_{\k}} f^0_{\k} \simeq 
\big[\tau_{\rm MFL}(\xi_\k,T) \,\partial_{\xi_{\k}} f^0_{\k} \big]\; 
\v_\k    \nonumber \\  
&& \;\;\;\;\;\;\;\;\; +\big[\tau^2_{\rm MFL}(\xi_\k,T)\, 
\partial_{\xi_{\k}} f^0_{\k}\big]\; \tau^{-1}_p(\theta)\; 
\big(\v_\k \times \hat{z}\big). \label{lorenzk}
\end{eqnarray}

We can readily obtain the longitudinal thermal conductivity 
$\kappa_{xx}$ 
\begin{eqnarray}
\kappa_{xx} &\simeq& \int^{\infty}_{-\infty} d\xi {\xi^2\over T} \, 
\big(-\partial_{\xi_\k} f_\k^0\big) \, \tau_{\rm MFL}(\xi,T) 
\langle N(\theta)v^2_x(\theta)\rangle_{\rm FS} \nonumber \\
&\simeq&  \, 0.65 \, 
\langle N(\theta)v^2_x(\theta)\rangle_{\rm FS}.   \label{kappaxx} 
\end{eqnarray}
Equations (\ref{kappaxx}) and (\ref{sigmaxx}) imply the observation of 
the Wiedemann-Franz law. The Lorenz number
\begin{eqnarray}
L={e^2 \kappa_{xx} \over T\sigma_{xx}} \simeq 2.24
\end{eqnarray}
being independent of $T$, shown by the dashed line in Fig. \ref{lxy},
is smaller than the universal value $L_0=\pi^2/3$.

\vskip 0.2cm
We insert Eq. (\ref{lorenzk}) into Eqs. (\ref{g2k}) and 
(\ref{heatcurrent}) to find the thermal Hall conductivity $\kappa_{xy}$
\begin{eqnarray}
&& \kappa_{xy} = {eB\over mc}\, \int^{\infty}_{-\infty} d\xi \, 
{\xi^2\over T} \, \big(-\partial_{\xi_\k} f_\k^0 \big) 
\bigg[ \bar{v}_2\; \tau^2_{\rm MFL}(\xi,T)  \nonumber \\
&& \;\;\;\;\;\;\;\;\;\;\;\;\;\;\;\;\;\;\;\;\;
+ \bar{v}_{\rm imp}\; 
\tau^3_{\rm MFL}(\xi,T) \bigg]    \\
&& \simeq  {eB\over m c} \bigg \{ \, 0.134 \, \lambda^{-2}_0 \, 
\bar{v}_2\;  \; T^{-1}+  0.029 \, \lambda^{-3}_0 \, \bar{v}_{imp} \; 
T^{-2} \bigg\}.    \label{kappaxy}
\end{eqnarray}
So $a_h$ and $b^{({\rm imp})}_h$ in Eq. (\ref{hallk}) become
\begin{eqnarray}
&& a_h={eB\over mc} \, 0.134 \, \lambda^{-2} \, \bar{v}_2
= 1.58 \,a_c,       \nonumber \\
&& b_h^{({\rm imp})}= {eB\over mc} \, 0.029 \, \lambda^{-3} \, 
\bar{v}_{imp} = 1.16 \, b_c^{({\rm imp})}, \label{abh}
\end{eqnarray}
which suggests the ratio $b_h^{({\rm imp})}/a_h \simeq 1.4 \, 
b_c^{({\rm imp})}/a_c$. Equations (\ref{sigmaxy}) and (\ref{kappaxy}) 
immediately lead to   
\begin{eqnarray}
L_{xy}={0.134 \, t+0.029 \over 0.085 \, t+ 0.025}\, , \label{lorenzkxy}
\end{eqnarray}
where $t=T \lambda_0 \bar{v}_2/\bar{v}_{\rm imp}$. $L_{xy}$ is a weak 
function of $T$, which is shown by the solid line in Fig.~\ref{lxy}. 
Therefore, the Hall-Wiedemann-Franz law is approximately obeyed.

\section{Discussion}
\begin{figure}[h]
\begin{picture}(250,200)
\leavevmode\centering\includegraphics{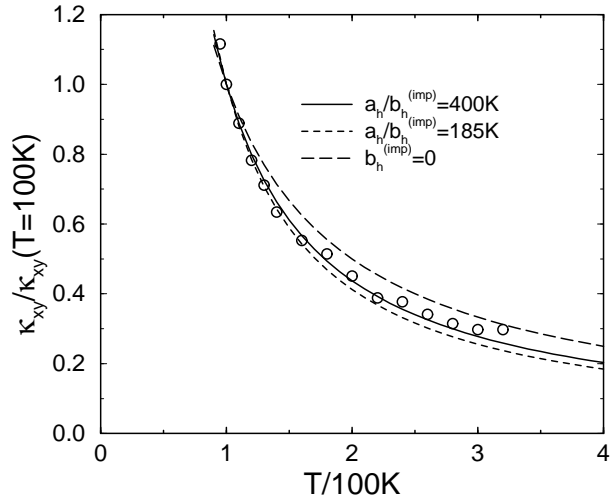}
\end{picture}
\caption{$\kappa_{xy}$ as a function of $T$. Open circles represent
experimental data quoted from Ref.\protect\cite{ong2}.}
\label{kxy1}
\end{figure}

The above result seems to contradict the experimentally observed 
violation of the Hall-Wiedemann-Franz law \cite{ong2}. It was already 
mentioned in Sec.~I, however, that $\kappa_{xy}$ and 
$\sigma_{xy}$ were measured in two different YBaCuO samples 
\cite{ong1,ong2}, which usually involve different impurity densities. 
In particular, the sample used in $\kappa_{xy}$ measurement \cite{ong2} 
is of a higher doping level, equivalent to the lower oxygen deficit in the 
CuO chains. Taking the oxygen vacancies in the CuO chains 
to be the out-of-plane impurities, we expect a weaker impurity scattering 
in this sample. It is therefore likely that with the {\it smaller} 
$\bar{v}_{\rm imp}$, $\kappa_{xy}$ shows a $T$ dependence closer to 
that of the customary term. But the important fact lies in the 
{\it coexistence} of these two terms. In Fig. \ref{kxy1}, we show the 
theoretical results of $\kappa_{xy}$ for various $a_h/b_h^{({\rm imp})}$, 
as compared with the experimental data (open circles). The theoretical 
curve representing $a_h/b_h^{({\rm imp})} = 400 $K fits the experiments 
very well. We speculate that this corresponds to the impurity scattering 
strength in the slightly overdoped YBaCuO sample used for the 
$\kappa_{xy}$ measurement. It is clear that to confirm this speculation 
a systematic experimental work is needed. An immediate suggestion is 
to examine $\sigma_{xy}$ in the slightly overdoped YBaCuO sample, in 
which $\kappa_{xy}$ was measured, to see a possible customary behavior 
of $\sigma_{xy}$.

\vskip 0.2cm
We have shown that both $\sigma_{xy}$ and $\kappa_{xy}$ are determined 
by a competition between the customary term and the unusual impurity 
term, which depends on the detailed band structure and the out-of-plane 
impurity scattering. There may also be an effect from the renormalization 
of the mass, the density of state, and the group velocity of the 
electrons by the coupling to 
the MFL fluctuations. However, as in the Fermi-liquid case, this effect 
cancels out in $\sigma_{xy}$ and $\kappa_{xy}$, leaving behind 
insignificant numerical prefactors of both the customary and unusual 
impurity terms. So we do not expect any qualitative change to the 
conclusion on the weak-$T$ dependence of $L_{xy}$ in the {\it same} sample.

\vskip 0.2cm
Finally, we address the effect, missing in the above discussion, of 
experimental boundary conditions in thermal conductivity measurements of 
no electric currents flowing through the samples. Imposing such a 
boundary condition causes charge accumulation at the sample edges, 
generating an effective electric field on the electronic heat carriers. 
As a result, both the thermal and thermal Hall conductivities acquire 
correction terms
\begin{eqnarray}
\tilde{\kappa}_{xx}& = &\kappa_{xx}-T\,S^2\sigma_{xx},
\label{tildekxx}\\ 
\tilde{\kappa}_{xy}&=& \kappa_{xy} -  2T \,S\, S_{xy}\,\sigma_{xx}
+ T\, S^2 \, \sigma_{xy},  \label{tildekxy}
\end{eqnarray}
where $S$ and $S_{xy}$ are the thermoelectric power and the Nernst 
coefficient 
multiplied by the magnetic field, respectively. $S$ and $S_{xy}$ have 
been studied independently in experiments \cite{varoy,Nernst}. 
Theoretically, they both require an odd-energy dependence of some 
combinations of $\v_\k$, $N(\theta)$, and the inelastic scattering rate, 
and therefore vanish to the first-order approximation in the present 
model. Exploring the temperature dependence of $S$ and $S_{xy}$ by 
including higher-order terms in energy is beyond the scope of this 
paper. We instead take experimental data for $S$ and $S_{xy}$ in the 
optimally doped high-$T_c$ cuprates to examine the effect of the 
correction terms in Eqs. (\ref{tildekxx}) and (\ref{tildekxy}). At 
$T=150$ K, typical experimental values are $\kappa_{xx}\sim 6 $W/mK, 
\cite{ando} $\sigma_{xx}\sim 10^6/\Omega$m, \cite{ando}
$S\sim 10^{-5}$ V/K, \cite{varoy} $\kappa_{xy}\sim 0.015 $ W/mK, 
\cite{ong2} $\sigma_{xy}\sim 10^4/\Omega$ m, \cite{ong1} and 
$S_{xy}\sim -3\times 10^{-9} $V/K. \cite{Nernst} Thus the correction 
term in Eq. (\ref{tildekxx}) is roughly $0.015$W/mK being two orders 
smaller than the first one. The two correction terms in Eq. 
(\ref{tildekxy}) turn out to be $10^{-5}$ W/mK and $1.5\times 
10^{-4}$ W/mK, respectively. Neglecting the second term on the right-hand
side of Eq. (\ref{tildekxy}), we set the temperature dependence of 
$S(T)=S_0-aT (a>0)$ (Ref. \cite{varoy}) in the third term, and find that 
it increases the impurity $T^{-2}$ term in $\kappa_{xy}$ slightly more 
than the customary $T^{-1}$ term. Therefore, our conclusion that 
$\kappa_{xy}$ and $\sigma_{xy}$ approximately obey the Wiedemann-Franz 
law in the {\it same} sample still holds.

\section{Conclusion}
In summary, we have examined the thermal Hall conductivity for 
MFL quasiparticles being scattered from the out-of-plane impurities, 
by following the phenomenological work on electrical Hall conductivity 
in Ref. \cite{va1}. We find that these impurities have very similar 
effects on both $\sigma_{xy}$ and $\kappa_{xy}$, and the effective 
Lorentz force they impose on the MFL quasiparticles depends weakly 
on the quasiparticle energy. This immediately leads to nearly the same 
temperature dependence of $\kappa_{xy}/T$ and $\sigma_{xy}$. We argue 
that the different temperature behaviors of $\kappa_{xy}/T$ and 
$\sigma_{xy}$ observed in the experiments may originate from different 
samples being measured. In measuring $\kappa_{xy}$ a slightly overdoped 
YBaCuO sample was used which has fewer out-of-plane impurities and hence 
a dominant customary behavior of $\kappa_{xy}$ \cite{ong2}. We suggest 
experiments to reexamine $\sigma_{xy}$ in this sample in order to 
see the same scaling of $\sigma_{xy}$ and $\kappa_{xy}/T$.

\vspace{0.2cm}
\begin{acknowledgments} 
We are very grateful to C. M. Varma for suggesting the present 
investigation and for useful discussions. We also acknowledge 
the kind hospitality of Bell Laboratories, where this work 
started.
\end{acknowledgments}

{\it Note added}: Recently, we noted the appearance of
an Erratum \cite{va3} for Ref. \cite{va1}. However, taking into account 
this Erratum does not change the results reported in the current work.

\end{document}